\documentclass[a4paper]{article}

\usepackage{INTERSPEECH2021}
\usepackage{bbm}
\usepackage{xcolor}
\title{Supervised Contrastive Learning for Accented Speech Recognition}
\name{Tao Han$^{\star}$, Hantao Huang$^{\star}$, Ziang Yang, Wei Han \thanks{$^{\star}$ indicates equal contribution.}}
\address{
  Mediatek Singapore}
\email{\{Taoh.Han, Hantao.Huang, ziang-yz.yang, Wei.Han\}@mediatek.com}

\begin{document}

\maketitle
\begin{abstract}
Neural network based speech recognition systems suffer from performance degradation due to accented speech, especially unfamiliar accents. In this paper, we study the supervised contrastive learning framework for accented speech recognition. To build different views (similar "positive" data samples) for contrastive learning, three data augmentation techniques including noise injection, spectrogram augmentation and TTS-same-sentence generation are further investigated. From the experiments on the Common Voice dataset, we have shown that contrastive learning helps to build data-augmentation invariant and pronunciation invariant representations, which significantly outperforms traditional joint training methods in both zero-shot and full-shot settings. Experiments show that contrastive learning can improve accuracy by 3.66\% (zero-shot) and 3.78\% (full-shot) on average, comparing to the joint training method.
\end{abstract}

\section{Introduction}

Recently, neural network based automatic speech recognition (ASR) has achieved impressive progress for real life applications.
However, unfamiliar or unseen accented speech, which is known to be the primary reason of unstable ASR system, often leads to unacceptable performance \cite{jain2018improved,viglino2019end}.

Many recent works are proposed to improve accented ASR. \cite{jain2018improved} built  a multi-task framework to explicitly supervise a multi-accent acoustic model and jointly train an accent classifier. \cite{winata2020learning} adopted Model-Agnostic Meta Learning (MAML) for accented ASR. It aims to provide a pre-trained model with better initialization to help the model to adapt new data distribution quickly \cite{DBLP:journals/corr/FinnAL17}. Although these methods work reasonably well, they require additional network architecture or complex training methodology. On the other hand, supervised contrastive learning only needs an additional loss function to learn robust representation \cite{chen2020simple,khosla2020supervised, saunshi2019theoretical}.  
It leverages the availability of "positive" and "negative" data samples and forces a higher and lower similarity between "positive" and "negative" samples, respectively. Most of the contrastive learning works focus on computer vision domain. In this work, we share similar spirit with \cite{khosla2020supervised} and bring supervised contrastive learning method to ASR domain.

We study in detail the effectiveness of supervised contrastive learning on accented ASR task. The main novelty of this work are, first, proposing supervised contrastive learning loss in ASR domain at the first time. Second, using data augmentation methods to create different views for supervised contrastive learning in ASR domain. This is especially important in pre-training stage to provide theoretically unlimited "positive" and "negative" samples. Third, we evaluate the performance benefits of each data augmentation methods under our proposed learning framework.  

Furthermore, we show the effectiveness of our proposed method by plenty experiments using Common Voice dataset \cite{ardila2019common}, which show that the proposed loss achieves better performance even without data augmentation methods. It improves the WER by 1.35\% and 0.7\% comparing to joint training method for zero- and full-shot, respectively. Applying data augmentation methods further enlarge the this improvement to 3.66\% and 3.78\%.

The broader impact of this work is that, the proposed contrastive learning framework can be applied to any character-based end-to-end ASR model to improve the robustness. In this paper, we use accented speech as an example to show the improvement of robustness. The proposed method can also be used to handle many other problems such as noise and reverberation problem, which will be our future works.  

\section{Related work}

\label{sec:related work}
\subsection{Contrastive learning}

Contrastive learning is an effective training method to learn robust representations in the supervised or unsupervised setting \cite{khosla2020supervised,saunshi2019theoretical}. 
The main idea is using different methods to create positive pairs and negative pairs. Target model is forced to increase and decrease the similarity between positive and negative pairs, respectively. Unsupervised contrastive learning benefits from stronger data augmentation but may lead to worse representation due to false negatives \cite{chen2020simple,khosla2020supervised}. It is due to automatically selected negative pairs could actually belong to the same semantic category, creating false negatives. For character-based or phoneme-based ASR systems, it is even more difficult to automatically select negative pairs due to small amount of text labels such as 26 characters. On the other hand, comparing to unsupervised contrastive learning, supervised contrastive learning leverages the label information to solve the false negatives problem and build robust representation \cite{khosla2020supervised}.  Supervised contrastive learning is thereby adopted in this work.

Among all contrasitve learning works, SimCLR is one of the most well known frameworks \cite{chen2020simple}. It uses siamese network to generate pair representations and simplifies previous contrastive learning algorithms without needing specific architecture or a memory bank. SimCLRv2 \cite{chen2020big} is further proposed to fully leverage the power of pretraining. We use similar framework used in SimCLR series and improve it in the speech domain.

Contrastive learning for speech applications is relatively new and needs more investigations. Contrastive predictive coding (CPC) is the first work utilizing contrastive learning. It uses a specially designed architecture to predict future segments of speech based on the past \cite{kharitonov2020data} in the pre-train stage. The main difference between our work and CPC is the definition of positive and negative views. CPC treats consecutive audio segments as positive views. Instead, we treat same target letter, such as the letter ``l"  in different accents or words, as positive views. This definition helps us learn a robust and accent invariant feature without changing of original ASR architecture. Furthermore, we apply different speech data augmentation techniques, such as spectrogram augmentation, and evaluate their effectiveness in the proposed contrastive learning method.

\subsection{Accented speech recognition}
To overcome the accent variability, neural network based systems typically adopt accent-dependent output layers similar to multilingual training of deep neural networks \cite{chen2015improving,huang2014multi}. This extends to a multi-task learning architecture to jointly learn an accent classifier and an acoustic model. For example, \cite{yang2018joint} proposed to jointly train a multi-accent acoustic model with accent identification network. Aside from joint training, \cite{winata2020learning} uses MAML, one of the most popular meta-learning methods, to learn a better initialization for fast adaptation to different data distribution. Different from normal training target, MAML targets to train model parameter to optimize the performance on unseen data distribution \cite{pmlr-v70-finn17a}. Another strategy is to learn accent-invariant features. \cite{sun2018domain} proposed a domain adversarial training algorithm to extract accent-invariant features from all accents. Our work shares the same spirit, but uses contrastive learning instead to extract the accent-invariant features to help recognize speech with unfamiliar accents.

\section{Methodology}
\label{sec:methodology}



\subsection{ASR model}
In general, an ASR model predicts characters, word-pieces or phonemes given input audio sequence. It can be considered as a classification problem that tries to classify every time-step input audio segments into different text labels. However, it is far more complicated than the classic classification problems due to its time dependency and large variance between different speakers under different environments. 

Without loss of generality, in this paper, a character based seq2seq ASR system is considered, which is trained by minimizing the following loss function:
\begin{equation}
    \mathcal{L}_{ASR} = -\sum_{i}{\log}P(y_i|x,y^*_{0:i-1};\theta),
\end{equation}
where $y_i$ represents current predicted label, $y^*_{0:i-1}$ is the ground truth before current time step, x represents input audio segments and $\theta$ represents model parameters.  
  
\begin{figure}
    \centering
    \includegraphics[width=0.4\textwidth]{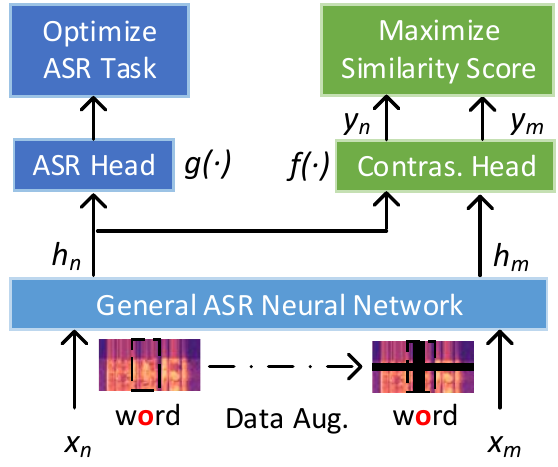}
    \caption{Contrastive learning framework for general ASR (End-to-end speech recognition) }
    \label{fig:framwork}
\end{figure}

\subsection{Contrastive learning framework for ASR}
Contrastive learning has become very popular recently especially in computer vision domain. It attempts to extract better image representation by maximizing agreement between different augmented views of the same figure using contrastive loss \cite{chen2020simple}. We follow the motivation of those algorithms and propose using contrastive learning in ASR models to boost both robustness and accuracy. 

Figure \ref{fig:framwork} depicts the proposed framework for one sample $x_n$ and its positive partner $x_m$. First, we convert original data $x_n$ to $x_m$ through data augmentation methods including noise injection, spectrogram augmentation and TTS-same-sentence generation. Both data are fed into an ASR neural network. Representations $h_n$ and $h_m$ are generated as outputs of the ASR. $h_n$ is then going through ASR head $g(\cdot)$, which  converts $h_n$ to the output logit of ASR task. The output dimension is 33 which represents 26 letters and 7 other labels. Furthermore, $h_n$ and $h_m$ will go through contrastive learning head $f(\cdot)$ and outputs $y_n$ and $y_m$ are regarded as all-invariant-embedding-space representation, which will be the input of contrastive learning loss to maximize their agreement. We choose feature size 16 ($1/32$ of the transformer hidden dimension 512) for this representation. This is because we experimentally find larger feature size shows no performance benefits. In this work, we use simple linear transfer function for both $g(\cdot)$ and $f(\cdot)$.

\begin{figure}
    \centering
    \includegraphics[width=0.47\textwidth]{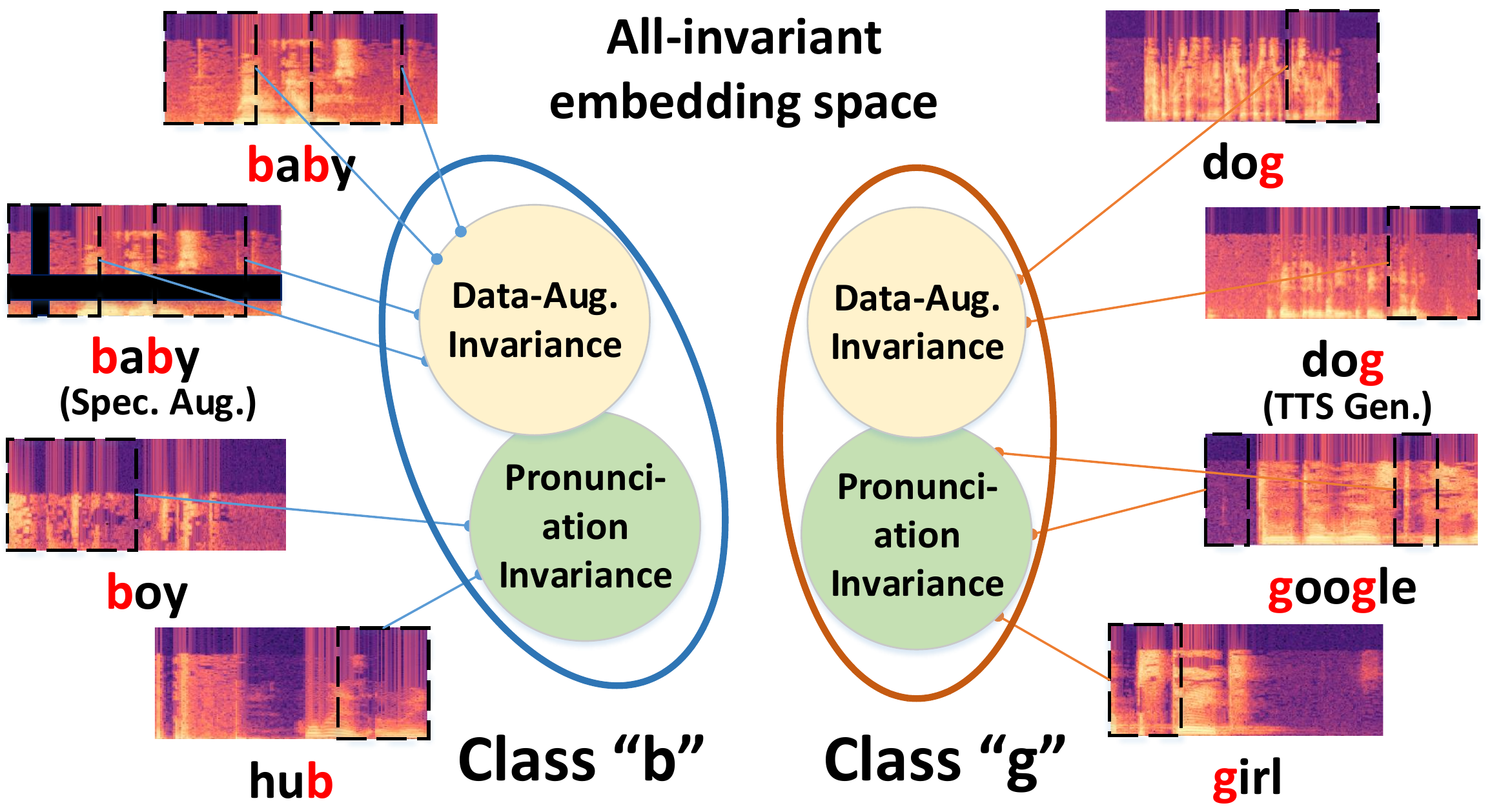}
    \caption{Class "b" and "g" all-invariant embedding space (For illustration purpose, data-augmentation invariant and pronunciation invariant representations will cluster together in real embedding space.)}
    \label{fig:all_inveriant}
\end{figure}

The total loss function is defined as
\begin{equation}
\label{asr_loss}
    \mathcal{L}_{ALL} = \mathcal{L}_{ASR} + \alpha \mathcal{L}_{con}.
\end{equation}
Note that we use $\alpha=1$ in the pre-train stage and $\alpha=0$ in the fine-tune stage. We find this leads to faster model convergence than the SimCLR framework \cite{chen2020simple}, which does not use the ASR loss $\mathcal{L}_{ASR}$ during the pre-train stage. 

$N$ represents the total number of characters in one batch.
With data augmentation, the total number could reach $2N$. Here, we use $n$ and $m$ to represent a positive pair of examples.
Positive pairs could be found from a given transcript such as: two "p" from the same word "happy"; two "b" from different words "boy" and "hub"; two "b" from the word "boy" and its data-augmented pair, etc.
The contrastive loss for positive example pairs $(n,m)$ is defined as
\begin{equation}
    \mathcal{L}_{con-pair}(n,m)=-{\log}\frac{{\exp}(sim(y_n,y_m)/\tau)}{\sum_{k}^{2N}\mathbbm{1}_{[k\neq{n}]} {\exp}(sim(y_n,y_k)/\tau)},
\end{equation}
where $\mathbbm{1}_{[k\neq{i}]}$ is an indicator function. It returns 1 when $k\neq{n}$, else 0. 
The similarity between two examples is represented by $sim(a,b)=\frac{ab^T}{\|a\|\|b\|}$ and $\tau$ is a scalar temperature parameter, which is set to 0.07 based on experiments. Here, we treat $y_m$ as a view of $y_n$. By adding more data augmentation techniques, we add more views of positive pairs as well as negative pairs; all the pair losses $\mathcal{L}_{con-pair}(n,m)$ are averaged to the final contrastive loss $\mathcal{L}_{con}$.

 As shown in Figure \ref{fig:all_inveriant}, our contrastive learning algorithm tries to map character representations to an all-invariant embedding space. It includes data-augmentation invariance and pronunciation invariance. Same character with different data augmentation methods is called data-augmentation invariance and same character in different words is regarded as pronunciation invariance. In general, same characters form positive group and different characters become negative pairs. Same group representations are optimized to be as close as possible and those from  different groups are pushed apart. As such, the all-invariant representation helps to build a robust and accent-invariant speech recognition system.



\section{Experiments}
\label{sec:experiments}

\subsection{Dataset}

In this work, we use the Mozilla collected dataset Common Voice Dataset \cite{ardila2019common}. 
More specifically, we use the same English dataset as \cite{winata2020learning} and adopt the cross-region performance  evaluation. 
The training dataset is comprised of five accents: Australia (22.86 hrs), England (64.19 hrs), Ireland (3.71 hrs), New Zealand (7.06 hrs) and United States (163.89 hrs), totally 261.78 hours and the validate dataset is based on Canada (20.20 hrs), Scotland (5.08 hrs) and South Atlantic (0.23 hrs).  The test dataset comprises of Africa (5.04 hrs), Hong Kong (1.21 hrs), Indian (29.09 hrs), Philippines (1.68 hrs) and Singapore (1.00 hrs). Note that the test dataset can be regarded as unseen or unfamiliar accent compared to training dataset.

\subsection{Experiment set-up}
\label{exp setup}
To have a fair comparison with our baseline using MAML \cite{winata2020learning}, we adopt the same transformer network with two encoder and four decoder layers. The 6-layer VGG-like CNN module \cite{simonyan2014very} is used to extract information from raw audio. The transformer model is built up with 512 dimension for all layers and has around 10.2M parameters.
We perform a two-stage training strategy for the supervised contrastive learning. The pre-train stage uses both contrastive loss and ASR loss as described in Equation \ref{asr_loss} with learning rate $2.83 \times 10^{-4}$, for $\alpha=1$ and the fine-tune stage uses learning rate $8 \times 10^{-5}$, for $\alpha =0$.  
For the zero-shot setting, the fine-tune stage is performed on the original training dataset (au, en, ir, nz, us) whereas for the full-shot setting, the fine-tune is further performed on the respective dataset (af, hk, in, ph, sg). Note that 100 samples from test dataset are randomly selected for testing, and remaining samples are used for full-shot fine-tune. 
We evaluate our model using beam search algorithm with below score function : 
 \begin{equation}
     \sum_{i}{\log}P(y_i|x,\widehat{y}_{0:i-1};\theta)+\lambda\sqrt{wc(\widehat{y}_{0:i-1})},  
 \end{equation}
where $\lambda = 0.1$, $beam size = 5$ and $wc$ represents word count.

We have two baselines: MAML \cite{winata2020learning} and joint training. For MAML, we refer to the best accuracy stated in \cite{winata2020learning}. It is trained with 200k iterations for pre-train and then perform fine-tune. For joint training, we use early stop mechanism with learning rate $2.83\times 10^{-4}$ during pre-train and $8 \times 10^{-5}$ during fine-tune stage, which is the same as our contrasitve-learning setting. Total iterations can reach around 800K. 

\section{Results and discussion}

\subsection{Effectiveness of supervised contrastive learning}

In this section, we demonstrate the effectiveness of the proposed methodology on accented speech recognition. Table \ref{tbl:effect} summarizes the performance of joint training, contrastive learning framework SimCLR proposed in \cite{chen2020simple} and proposed contrastive learning method. The main difference between SimCLR and our method is that in pre-train stage, our method uses both ASR loss $\mathcal{L}_{ASR}$ and contrastive loss $\mathcal{L}_{con}$, while the SimCLR only uses $\mathcal{L}_{con}$. It can be observed that SimCLR even gets worse performance than joint training. This is mainly due to the difficulty of learning accent-invariant representations directly without the help of ASR loss in pre-train stage. However, by adding ASR loss, as discussed in Equation \ref{asr_loss}, along with contrastive loss during pre-train, WER improves from 46.35\% to 44.85\%. This indicates that contrastive learning task and ASR task benefit each other and lead to the better convergence.  
\begin{table}[!hb]
\small
\centering
\caption{Zero-shot performance comparison on WER (\%) \label{tbl:effect}}
\scalebox{0.92}{
\begin{tabular}{|c|c|c|c|c|c|c|}
\hline
Mehods                                                          & Af    & HK    & Ind   & Phip  & Sg    & Avg. \\ \hline
Joint train                                                 & 40.88 & 44.90 & 63.73 & 43.77 & 37.71 & 46.20   \\ \hline
\begin{tabular}[c]{@{}c@{}}SimCLR \cite{chen2020simple} \end{tabular} & 40.44 & 47.70 & 66.36 & 42.08 & 35.18 & 46.35   \\ \hline
\begin{tabular}[c]{@{}c@{}}Contras.\\ Proposed\end{tabular}     & 38.57 & 44.25 & 64.69 & 40.99 & 35.76 & 44.85   \\ \hline
\end{tabular}
}
\end{table}

Table {\ref{tbl:cmp}} compares our methods with joint training and MAML in details and shows the effectiveness of different data augmentation methods. 
Comparing two baselines, it is supervised that the joint training model outperforms MAML in both zero-shot and full-shot settings. This may due to the training strategy and total iteration number difference mentioned in Section \ref{exp setup}. Our proposed training methodology outperforms the joint model by 1.35\% in zero-shot setting and 0.7\% in full-shot setting even without data augmentation. This is because self-contrastive behaviour improves pronunciation invariance regardless of existence of data augmentation as shown in Figure \ref{fig:all_inveriant}. 

\begin{table*}
\centering
\caption{Accented speech recognition accuracy under zero-shot and full-shot configurations on WER (\%) \label{tbl:cmp}}
\small
\scalebox{0.91}{
\begin{tabular}{|c|c|c|c|c|c|c|c|c|c|c|c|}
\hline
\textbf{Accents} & \textbf{MAML \cite{winata2020learning}} & \textbf{Joint} & \textbf{\begin{tabular}[c]{@{}c@{}}Joint\\ + Noise\end{tabular}} & \textbf{\begin{tabular}[c]{@{}c@{}}Joint \\ + Aug\end{tabular}} & \textbf{\begin{tabular}[c]{@{}c@{}}Joint \\ + TTS\end{tabular}} & \textbf{\begin{tabular}[c]{@{}c@{}}Joint\\ + ALL\end{tabular}} & \textbf{Contras.} & \textbf{\begin{tabular}[c]{@{}c@{}}Contras.\\  + Noise\end{tabular}} & \textbf{\begin{tabular}[c]{@{}c@{}}Contras.\\ + Aug\end{tabular}} & \textbf{\begin{tabular}[c]{@{}c@{}}Contras.\\ + TTS\end{tabular}} & \textbf{\begin{tabular}[c]{@{}c@{}}Contras.\\ + ALL\end{tabular}} \\ \hline
\multicolumn{12}{|c|}{Zero shot (only use pre-train data for training)}                                                                                                                                                                                                                                                                                                                                                                                                                                                                                                                                                                       \\ \hline
\textbf{Af}      & 39.27                 & 40.88          & 38.24                                                            & 34.62                                                           & 39.56                                                           & 34.62                                                          & 38.57             & 36.59                                                                & 31.76                                                             & 38.02                                                             & \textbf{30.44}                                                    \\ \hline
\textbf{HK}      & 41.30                 & 44.90          & 41.71                                                            & 41.58                                                           & 45.66                                                           & 42.86                                                          & 44.25             & 41.71                                                                & 38.39                                                             & 44.01                                                             & \textbf{34.31}                                                    \\ \hline
\textbf{Ind}     & 61.17                 & 63.73          & 64.52                                                            & 60.71                                                           & 65.83                                                           & 57.42                                                          & 64.69             & 61.76                                                                & \textbf{53.09}                                                    & 65.31                                                             & 56.50                                                             \\ \hline
\textbf{Phip}    & 49.32                 & 43.77          & 43.53                                                            & 37.48                                                           & 40.51                                                           & 34.95                                                          & 40.99             & 41.23                                                                & 33.86                                                             & 39.42                                                             & \textbf{30.71}                                                    \\ \hline
\textbf{Sg}      & 54.90                 & 37.71          & 41.21                                                            & 33.72                                                           & 40.33                                                           & 32.75                                                          & 35.76             & 34.89                                                                & \textbf{30.22}                                                    & 36.83                                                             & 32.36                                                             \\ \hline
\textbf{Avg.}    & 49.19                 & 46.20          & 45.84                                                            & 41.62                                                           & 46.38                                                           & 40.52                                                          & 44.85             & 43.24                                                                & 37.46                                                             & 44.72                                                             & \textbf{36.86}                                                    \\ \hline
\multicolumn{12}{|c|}{Full shots (use task specific training data for training)}                                                                                                                                                                                                                                                                                                                                                                                                                                                                                                                                                           \\ \hline
\textbf{Af}      & 33.63                 & 36.37          & 35.82                                                            & 33.08                                                           & 37.69                                                           & 31.65                                                          & 36.37             & 36.15                                                                & 29.34                                                             & 33.52                                                             & \textbf{27.91}                                                    \\ \hline
\textbf{HK}      & 34.90                 & 43.37          & 43.24                                                            & 37.37                                                           & 40.43                                                           & 38.78                                                          & 44.13             & 41.33                                                                & 36.99                                                             & 42.47                                                             & \textbf{33.80}                                                    \\ \hline
\textbf{Ind}     & 46.85                 & 45.47          & 41.79                                                            & 42.05                                                           & 43.89                                                           & 37.45                                                          & 45.60             & 42.18                                                                & 39.68                                                             & 42.05                                                             & \textbf{35.61}                                                    \\ \hline
\textbf{Phip}    & 44.23                 & 37.61          & 32.41                                                            & 27.33                                                           & 36.15                                                           & 30.83                                                          & 37.61             & 34.70                                                                & \textbf{25.63}                                                    & 32.89                                                             & 27.93                                                             \\ \hline
\textbf{Sg}      & 51.65                 & 39.16          & 37.22                                                            & 32.65                                                           & 38.19                                                           & 34.01                                                          & 34.79             & 32.56                                                                & 28.70                                                             & 32.94                                                             & \textbf{28.57}                                                    \\ \hline
\textbf{Avg.}    & 42.25                 & 40.40          & 38.10                                                            & 34.50                                                           & 39.27                                                           & 34.54                                                          & 39.70             & 37.38                                                                & 32.07                                                             & 36.77                                                             & \textbf{30.76}                                                    \\ \hline

\end{tabular}
}
\end{table*}

\begin{figure*}[!t]
    \centering
    \includegraphics[width=0.92\linewidth]{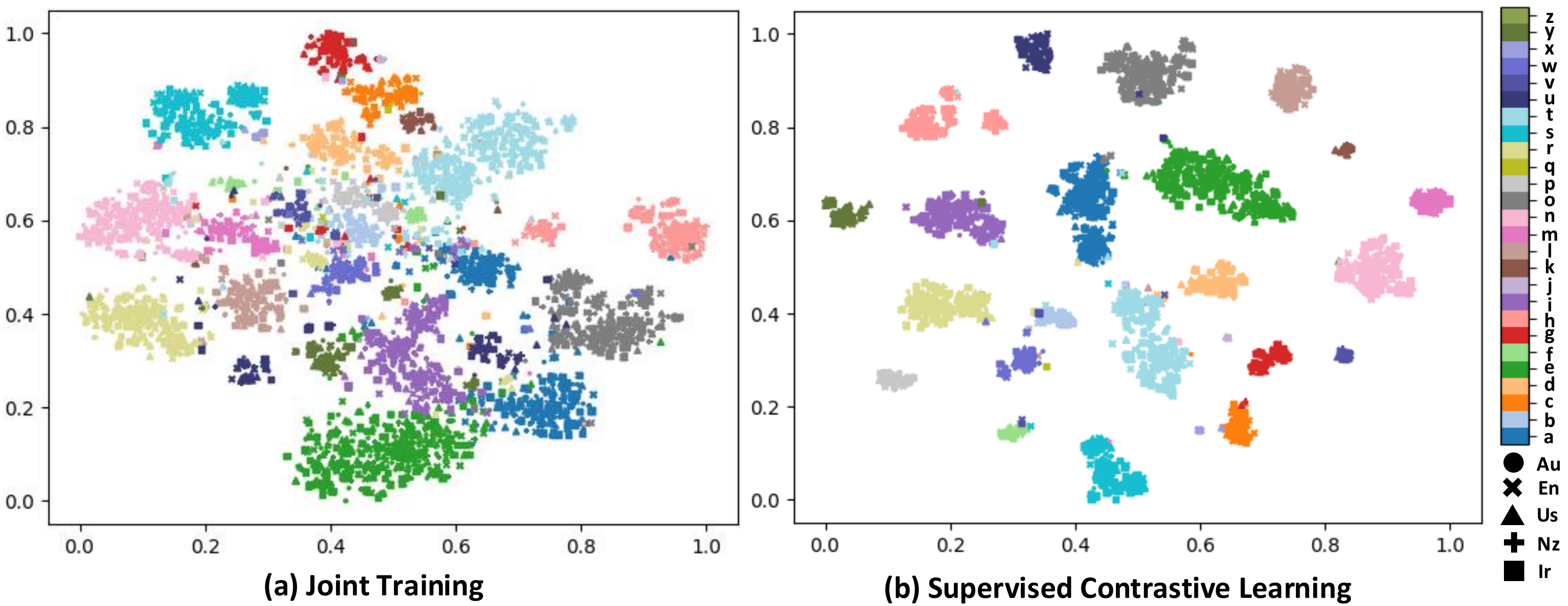}
    \caption{t-SNE visualization of 26 different characters for representation $h$ in Figure \ref{fig:framwork}}
    \label{fig:t-sne}
\end{figure*}

Apart from pronunciation invariance, data-augmentation invariance plays a vital role. Three data augmentation methods are adopted to extend the view of contrastive learning: 1) \textbf{Noise injection}: we inject noise data from \cite{6701894} with 50\% probability during training;
 2) \textbf{Spectrogram augmentation}: we augment the spectrogram based on \cite{park2019specaugment} with 25\% probability;
 3) \textbf{TTS-same-sentence generation}: the speech for the same sentence is generated using WaveGlow \cite{prenger2019waveglow} and Tacotron-V2 \cite{shen2018natural} (Total length: 191.94 hrs).   
 Table \ref{tbl:cmp} shows that adding data augmentation can benefit both joint training and contrastive learning. Among all the data augmentation methods, spectrogram augmentation is proved to be the best in our case. It improves WER by 7.39\% and 7.63\% absolutely for zero- and full-shot, respectively. As regards to TTS-same-sentence generation, it does not benefit the WER a lot in zero-shot case. It only improves the WER by 0.13\% absolutely. It may be due to the fact that the generated TTS sentences can be regarded as same-person speech. Adding 191.94 hours same-person speech into training dataset may cause the model tends to be speaker dependent. However, it does provide more views for contrastive learning and therefore, when more data is available, e.g., full-shot, it improves the WER by 2.93\% absolutely. The effectiveness of this method could be further improved if using more advanced TTS model that can generate sentences with different voices.
 
Overall, compared with joint training models, our proposed contrastive learning training methods gains more WER improvement by using data augmentation methods. The reason is that, for contrasitve learning, the data augmentation not only increases the training variance for ASR task but also adds more views which creates more "positive" and "negative" pairs. If we use all data augmentation methods, joint training WER improves 5.68\% (zero-shot) and 5.86\% (full-shot), whereas contrastive learning WER improves 7.99\% (zero-shot) and 8.94\% (full-shot). This demonstrates the importance of applying data augmentation in contrasitve learning methods.

\subsection{Embedding space visualization}
To further demonstrate the ability of our proposed method extracting accented invariance representations, we prepared Figure \ref{fig:t-sne} for t-SNE \cite{t-sne} visualization of 26 character representations for hidden representation $h$ in Figure \ref{fig:framwork}. 100 sentences are randomly selected from training dataset. It clearly shows that when using  contrastive learning method, points belonging to the same letter cluster together with a clear separation from other classes as shown in Figure \ref{fig:t-sne}(b), while in the joint training method, the separation from different letters is much smaller as shown in Figure \ref{fig:t-sne}(a). We have also shown that the same letters from different accents like Australia and England also cluster together, which further demonstrate the effectiveness of proposed contrastive learning loss to learn an accent-invariant representation. This visualization matches the intuition of contrastive learning in accented speech recognition.

\section{Conclusion}
\label{sec:conclusion}
In this paper, we introduce a supervised contrastive learning framework for a robust speech recognition system.
Contrastive-learning loss is applied to perform the pre-train and fine-tune without changing model architecture. Three data augmentation techniques (noise injection, spectrogram augmentation and TTS-same-sentence generation) are evaluated under the proposed framework and spectrogram augmentation benefits the most. Experiments on the Common Voice dataset show that contrastive learning with data augmentation can effectively improve the word error rate from 40.52\% to 36.86\% and 34.54\% to 30.76\% compared to the joint training under zero- and full-shot configuration, respectively.

\vfill\pagebreak
\vfill\pagebreak
\bibliographystyle{IEEEtran}
\bibliography{main}

\end{document}